\documentclass[aps,prd,twocolumn,nofootinbib,showpacs,amssymb,amsmath,floatfix]{revtex4}

\usepackage{graphicx}
\usepackage{epsfig}
\usepackage{subfigure}

\def \apjl{ApJ}

\def \apj{ApJ}

\def \mnras{MNRAS}

\def \nat{Nature\ }
\def\msun{{\,M_\odot}}

\def\alt{\raise0.3ex\hbox{$\;<$\kern-0.75em\raise-1.1ex\hbox{$\sim\;$}}}
\def\agt{\raise0.3ex\hbox{$\;>$\kern-0.75em\raise-1.1ex\hbox{$\sim\;$}}}

\newcommand{\bw}{\begin{widetext}}
\newcommand{\ew}{\end{widetext}}

\newcommand{\lsim}{\,\rlap{\raise 0.35ex\hbox{$<$}}{\lower 0.7ex\hbox{$\sim$}}\,}
\newcommand{\gsim}{\,\rlap{\raise 0.35ex\hbox{$>$}}{\lower 0.7ex\hbox{$\sim$}}\,}

\begin{document}

%

\title{The Fermi Bubbles: Giant, Multi-Billion-Year-Old Reservoirs of Galactic Center Cosmic Rays} 

\author{
  Roland M.~Crocker$^{1}$\footnote{Roland.Crocker@mpi-hd.mpg.de},
  Felix Aharonian$^{2,1}$\footnote{Felix.Aharonian@dias.ir}
}

\affiliation{
 $^1$Max-Planck-Institut f{\" u}r Kernphsik, P.O. Box 103980 Heidelberg, Germany \\
 $^2$Dublin Institute for Advanced Studies, 31 Fitzwilliam Place, Dublin 2, Ireland
}

\date{\today}

\begin{abstract}
Recently evidence has emerged for enormous features in the $\gamma$-ray sky observed by the Fermi-LAT instrument: 
bilateral `bubbles' of  emission centered on the core of the Galaxy and extending to  around  $\pm$10 kpc  above and below the Galactic plane.
These structures are coincident with a non-thermal microwave `haze' found  in WMAP data and an extended region of X-ray emission detected by ROSAT.
The bubbles' $\gamma$-ray emission is characterised by a 
hard and relatively uniform spectrum, relatively uniform intensity, and an overall luminosity $\sim 4 \times 10^{37}$ erg/s,
around one order of magnitude larger than their microwave luminosity
while more than order of magnitude less than their X-ray luminosity.
Here we show that the bubbles are naturally explained as due to a population of relic cosmic ray  protons and heavier ions  injected by processes associated with extremely long timescale ($\gsim8$ Gyr) and high areal density star-formation in the Galactic center.
\end{abstract}


\maketitle

A recent analysis  \cite{Dobler2009,Su2010}  of Fermi-LAT  \cite{Atwood2009} $\gamma$-ray data has revealed  
two, enormous bubble-like  emission features centered on the core of the Galaxy and extending to  around  $\pm$10 kpc  above and below the Galactic plane.
At lower Galactic latitudes these structures are coincident with a non-thermal microwave `haze' found  in WMAP 20-60 GHz data  \cite{Finkbeiner2004,Dobler2008} and an extended region of diffuse X-ray emission detected by ROSAT \cite{Snowden1997}.

\begin{figure}
\renewcommand \thefigure{1}
\centerline{\includegraphics[width=0.5\textwidth]{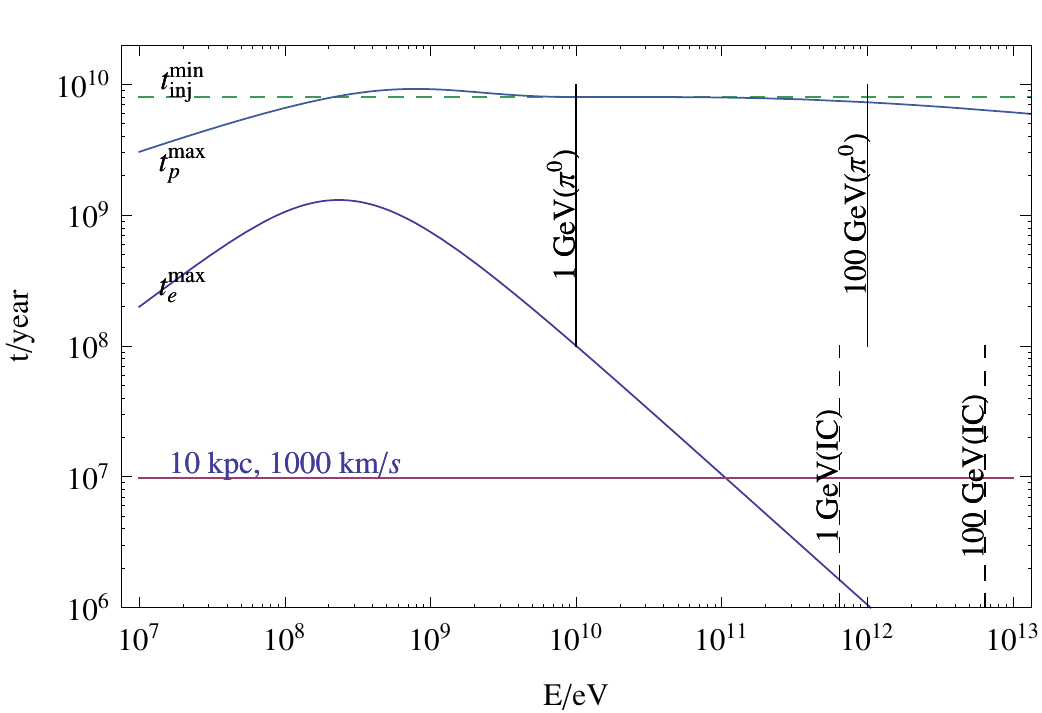}}
\caption{{\bf Timescales for CR protons and electrons in the bubble medium. }
Key:-  `$t_p^{max}$': maximum cooling time, 1/($d \ $ln $E/d$t)), 
for protons given ionization and $pp$ collisions in the $n_H  \simeq  0.005$ cm$^{-3}$ bubble plasma.
The vertical solid lines show the approximate mean proton primary energy for daughter $\gamma$-rays of 1 GeV and 100 GeV.
`$t_{inj}^{min}$': minimum formation time ($\sim$ 8 Gyr) for the bubbles in order that the system reach saturation.
`$t_e^{max}$': maximum timescale (adopting
 $U_B \ll U_{CMB}$ or $B \ll 3 \times 10^{-6}$ G) of electron cooling due to the combination of ionization and bremsstrahlung in the plasma and IC-scattering on the CMB.
The solid horizontal line shows the timescale for transport of electrons out to the full extent of the bubbles on a (very fast) 1000 km/s wind.
The vertical dashed lines show the energies required for an electron to IC scatter a CMB photon to 1 and 100 GeV.
Much higher energy electrons than protons are required to produce a daughter $\gamma$-ray of the same energy.
 %
}
\label{fig_Timescales}
\end{figure}

A natural explanation of these structures \cite{Dobler2009} would be that they are due to the same population of highly-relativistic ($\gsim 50$ GeV) cosmic ray (CR) electrons which  synchrotron radiate at multi-GHz frequencies and  simultaneously produce $\gsim$ 1 GeV 
$\gamma$-rays through the inverse Compton (IC) process.
However, given the severe radiative energy losses experienced by electrons, 
the hard spectrum, uniform intensity, vast extension, and energetics of the bubbles render the origin of this
particle population extremely mysterious (see fig.~\ref{fig_Timescales}) \cite{Finkbeiner2004,Dobler2008,Dobler2009,McQuinn2010,Su2010}.
In particular, transport of $\geq$TeV, IC-radiating electrons to the requisite distances from the plane would require velocities of $> 0.03 \ c$, too fast for a Galactic wind (this is a conservative lower limit as only electron energy losses on the 2.7 K CMB are accounted for).
If diffusive, a diffusion coefficient of $\sim 10^{31}$ cm$^2$/s would be required for $E_e = 1$ TeV, 1--2 orders of magnitude {\it larger} than the Galactic plane value.
One might postulate an in-situ electron acceleration/injection process to surmount these difficulties \cite{Su2010} but here
spectral considerations present a severe test. 
The $\sim E_\gamma^{-2}$ $\gamma$-ray spectrum might be due to IC emission from a cooled $\sim E_e^{-3}$ electron population but there is a robustly-detected \cite{Su2010} hardening in the $\gamma$-ray spectrum below $\sim$ 1 GeV.
In order to produce such a break either a {\it unique} (over the age of the Galaxy) injection event of age $\sim 10^6$ years or a sharp,
 $\sim$1 TeV low-energy hardening or cut-off in the {\it injection} spectrum of electrons is required.
The former seems unlikely as there are no indications of such an event occurring over this timescale in the Galactic center (GC).
The latter cannot be excluded in principle: an in-situ acceleration process (involving, e.g., stochastic acceleration or magnetic reconnection) that produced a spectrum harder than $\sim E_e^{-1}$ for $E_e \lesssim 1$ TeV would suffice, though such seems {\it ad hoc}.

In contrast to the difficulties presented by the electron/IC case, we show below that a CR {\it proton} population -- associated with extremely long timescale star-formation (SF) in the GC and injected into the bubbles by a wind-- can naturally explain the $\gamma$-ray structures {\it provided the protons are trapped for timescales approaching $10^{10}$ years}.
Our basic picture is that, granted this timescale, the majority of the power going into non-thermal protons is lost into $pp$ collisions on the bubbles' low-density plasma (itself the `wind-fluid' also injected by GC SF) and subsequently reprocessed into $\gamma$-rays, electrons, positrons and neutrinos.
Because the protons are trapped in our scenario their steady-state distribution mirrors their injection spectrum.
Requiring that the bubbles' hard-spectrum and overall (1-100 GeV) luminosity be reproduced implies a hard spectrum population of CR protons of (time-averaged) power $\sim 10^{39}$ erg/s  injected over multi-Gyr timescales.
Given that our independent \cite{Crocker2010} work suggests that i) the GC launches $\gtrsim 10^{39}$ erg/s in hard-spectrum CRs on a strong outflow, ii) the morphology of the bubbles clearly privileges the GC \cite{Su2010}, and iii) the GC is perhaps the single, spatially-localized site in the Galaxy where SF over multi-Gyr timescales is assured \cite{Serabyn1996,Figer2004} this SF ultimately offers a compelling explanation of the $\gamma$-ray and -- incidently -- microwave and X-ray phenomenology of the bubbles as we now explore.

Cosmic ray hadrons (in principle protons and heavier ions, but, for simplicity, we refer to the dominant protons below) 
undergo collisions with ambient matter creating daughter mesons (mostly pions), the neutral component of which decays into $\gamma$-rays.
This process  explains most of the diffuse $\gamma$-ray emission detected at $\gsim$100 MeV energies from the Galactic plane.
The hadronic $\gamma$-ray luminosity of a region scales linearly in its thermal (gas) and relativistic (CR hadron) populations, thus, 
assuming that high energy CRs are highly penetrating, it is generally taken that astrophysical markers of ambient gas density should trace hadronic $\gamma$-rays.
This heuristic does not always apply, however.
If the timescale for particle acceleration events in a system is smaller than all other (energy loss or escape) timescales the system will be in (quasi) steady-state.
If the timescale for particle energy loss (via hadronic collisions here) is less than the escape time the system qualifies as a `thick target'.
If both conditions pertain the system is in {\it saturation}.
The (hadronic) $\gamma$-ray luminosity of a region is $L_\gamma \simeq N_p/t_{pp\to\pi^0}$, where the $N_p$ is the region's steady-state proton population and $t_{pp\to\pi^0}$ is the timescale for neutral pion production in such collisions.
In saturation, however, $N_p \simeq \dot{Q_p} \ t_{pp}$ 
(because energy loss through $pp$ collisions is the dominant loss process)
and -- reflecting the almost equal production of $\pi^0, \pi^+$, and $\pi^-$ in hadronic collisions 
so that $t_{pp\to\pi^0} \simeq 3 \  t_{pp}$ -- we have that 
$L_\gamma \simeq  \dot{Q_p}/3$.
Thus, {\it in saturation about a third of the 
power injected into relativistic CRs will emerge in $\gamma$-rays (of all energies), independent of in-situ 
gas density, interaction volume, and CR injection time.}

\begin{figure}
\renewcommand \thefigure{2}
\centerline{\includegraphics[width=0.5\textwidth]{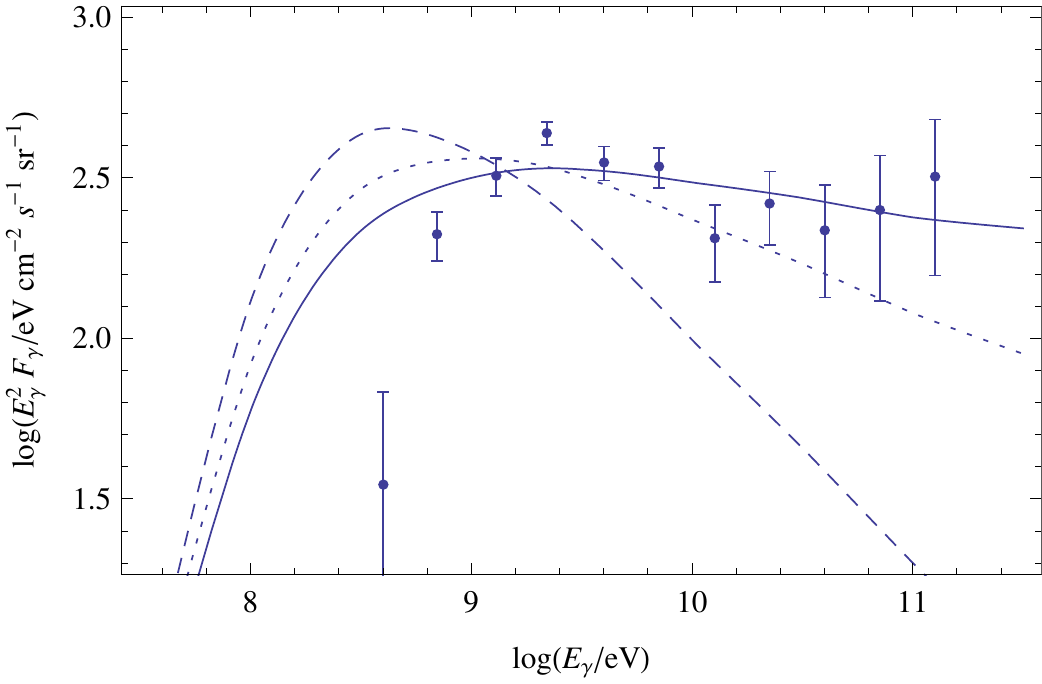}}
\caption{{\bf Fermi bubble $\gamma$-ray spectrum} \cite{Su2010}. The error bars are $1\sigma$. 
The solid curve shows the best-fit spectrum ($\propto E_p^{-2.1}$),
the dotted curve shows the steepest reasonable spectrum ($\propto E_p^{-2.3}$),
and, for comparison, the dashed curve is the spectrum expected were the bubbles suffused with protons having the 
$\propto E_p^{-2.7}$ distribution of the Galactic disk.}
\label{fig_plotBubbleSpectrum}
\end{figure}

Assuming saturation, a hadronic  scenario reproduces a number of the aspects of the phenomenology of the Fermi bubbles:
i) the reported \cite{Dobler2009,Su2010} {\it hard} $\gamma$-ray spectrum is explained;
consider the
 the situation in the Galactic plane.
 Here diffusive confinement of the CRs leads to a steepening of the steady state spectrum  to $\propto E^{-2.7}$.
In stark contrast,
{\it there is no energy-dependent confinement effect in the bubbles}.
So, given the almost energy-independent $pp$ loss time, 
we see the spectrum of the CRs {\it as injected at their acceleration sites} (evidently $\propto E_p^{-2.1}$) mirrored by the 
bubble $\gamma$-rays.
ii)
Decay kinematics, however, do imply a bump at low energy in the $\gamma$-ray spectrum around $E_\gamma = m_{\pi^0}/2 \simeq 70$ MeV.
This is mapped to a downturn below $\sim$GeV on a spectral energy distribution plot (see fig.~\ref{fig_plotBubbleSpectrum}; n.b.~the lowest and highest energy data points suffer from large systematic uncertainties \cite{Su2010} in the subtraction of back/foreground emission); 
this explains the afore-mentioned  down-break in the bubble $\gamma$-ray data.
iii) In saturation circumstances conspire to establish a constant volume emissivity -- regions of higher gas density necessarily have attenuated steady-state proton populations because of extra cooling by the gas.
This effect tends to give a uniform intensity of emission, as observed, despite the expected variations in target gas density.

To satisfy the requirement that the system be in saturation we require that the timescale associated with energy loss through 
$pp$ collisions (the dominant loss process at relevant energies) is less than the timescale over which CRs have been injected: 
\begin{equation}
t_{pp} \ \equiv \ 1/(\kappa_{pp} n_H \ \sigma_{pp} c) \ < \ t_{inj} \ .
\label{eqn_condn}
\end{equation} 
Given the inelasticity satisfies $\kappa_{pp} \simeq 0.5$, $\sigma_{pp} \simeq 5 \times 10^{-26}$ cm$^2$, and an  
 plasma density $\sim0.01$ cm$^{-3}$ \cite{Su2010},  this condition implies 
$ t_{inj} \ \gsim \ 5 \times 10^9 \ (n_H/0.01 \ {\rm cm}^{-3})^{-1}$ yr.
%
%

It is remarkable that the rate of SF in today's GC can supply the  {\it power and mass} required to explain the bubbles' $\gamma$-ray emission.
IRAS data \cite{Launhardt2002} imply that the inner $1.5^\circ$ (in diameter) around the GC emits a total infrared luminosity $L_{TIR} \simeq 1.6 \times 10^{42}$ erg/s which implies \cite{Kennicutt1998} a star formation rate for the region of $SFR \simeq 0.08  \ \msun$/yr.
%
 %
Such a $SFR$, in turn, implies \cite{Thompson2007} a supernova (SN) rate throughout the same region of 0.04/century (uncertain by  $\sim$2) leading to a total power injected by supernovae of $\dot{E}_{SN} \simeq 1.3 \times 10^{40} \ E_{51}$ erg/s, where $E_{51} \equiv E_{SN}/10^{51}$ erg is the mechanical energy released per SN.
Assuming \cite{Hillas2005} $\sim 10$\% of a SN's mechanical energy goes into accelerating non-thermal particles, this implies GC 
generates a power of $\sim 1.3 \times 10^{39}$ erg/s in CRs.

From Fermi data \cite{Meurer2009} we infer that  the $1^\circ \times 1^\circ$ field around the GC emits
 $\sim 3 \times 10^{36}$ erg/s s in  $ >1$  GeV $\gamma$-rays,
much less than the $\gsim 10^{38}$ erg/s expected were the system calorimetric and given its SN rate  
(the GC GeV emission is
substantially polluted by line-of-sight and point source emission in any case; 
the $\gamma$-ray flux at $\sim$ TeV energies reported by HESS  \cite{Aharonian2006} also indicates the system is far from calorimetric \cite {Crocker2010}). 
The GC, then, loses $\sim 10^{39}$ erg/s in hard-spectrum CRs into the Galactic environment.
In comparison, the bubbles emit $4 \times 10^{37}$ erg/s in 1 to 100 GeV $\gamma$-rays \cite{Su2010}, implying that CR protons in the energy range 10 to 1000 GeV lose $\sim 1.2 \times10^{38}$ erg/s in $pp$ collisions or $\sim 3.6 \times 10^{38}$ erg/s integrating the  
$\sim E_p^{-2}$ distribution from GeV to $10^6$ GeV.
Accounting for ionization losses by sub-relativistic protons and adiabatic energy losses at all energies, 
bubble protons lose a total $\sim10^{39}$ erg/s in steady state, precisely accounting 
for the CR power injected at the GC.

A prime candidate for a process that removes most of the GC's CRs is a `super-wind' 
driven by the same SF processes ultimately responsible for the CR acceleration.
Such winds are detected
emerging from the nuclei of many star-forming galaxies \cite{Strickland2009}.
A detailed accounting \cite{Crocker2010b}  -- taking into account mass loss from stars and injected by supernovae -- finds that the mass injected into this wind by the inner $1.5^\circ$ region is $\dot{M}_{\rm wind} \simeq (0.02-0.03) \msun$/year.
 The asymptotic speed of such a wind will scale as
 $v_{wind} \sim \sqrt{2 \ \dot{E}_{SN}/\dot{M}_{\rm wind}}$ which evaluates to 1200 km/s (for an adiabatic outflow), 
 larger than the  gravitational escape speed from the region ($\sim 1000$ km/s \cite{Muno2004}).
There are, however, both strong empirical \cite{Keeney2006} and theoretical \cite{Rodriguez-Gonzalez2009} indications that, given radiative losses, 
such an outflow should stall
at a height $\lsim$15 kpc, 
consistent with the
$\sim$10 kpc half-height of the bubbles.

With the above mass injection rate, we can re-visit the the minimum formation timescale for the bubbles under the assumption that the CR target gas is precisely that carried out of the GC on the putative wind.
The gas density in the bubbles is $n_H \simeq \dot{M}_{\rm bbl} \times t_{inj}/V_{\rm bbl}$ 
where $ \dot{M}_{\rm bbl}^{min} \leq \dot{M}_{\rm bbl} \leq \dot{M}_{\rm wind}$, given that the net mass growth of the bubbles has to allow for plasma cooling and falling back to the plane ($ \dot{M}_{\rm bbl}^{min} \equiv 0.004 \msun/yr$ corresponds to $t_{inj} = 13$ Gyr).
In concert with eq. \ref{eqn_condn} we find 
\begin{eqnarray}
t_{inj}  \gsim  \sqrt{\frac{m_p \ V_{\rm bbl}}{\kappa_{pp}  \sigma_{pp}  c \ \dot{M}_{\rm bbl}}}
\simeq 8 \times 10^9 \ \textrm{yr} \sqrt{\frac{\dot{M}_{\rm bbl}}{0.01 \msun/\textrm{yr}}}  \nonumber \ ,
\\
n_H \gsim  \sqrt{\frac{\dot{M}_{\rm bbl}}{m_p V_{\rm bbl} \kappa_{pp} \sigma_{pp} c }} 
\simeq  0.005 \ \textrm{cm}^{-3}  \sqrt{\frac{ \dot{M}_{\rm bbl}}{0.01 \msun/\textrm{yr}}} \ .
\end{eqnarray}
%

This gas is injected at the base of the wind dominantly in plasma form and will radiate an amount of power in thermal bremsstrahlung X-rays
controlled by the plasma density (already determined) and temperature of 
\begin{eqnarray}
L_{\rm plasma}  =  \Lambda(T) \ n_H^2 V_{\rm bbl} > L_{\rm plasma}^{\rm min} \ \equiv \ \Lambda(T) \ \frac{ \dot{M}_{\rm bbl}}{m_p \kappa_{pp} \sigma_{pp} c} \ \nonumber, 
\end{eqnarray}
where $ \Lambda(T)$ is the plasma cooling coefficient \cite{Raymond1976}
and the last inequality follows from the preceding equations.
In steady state, conservation of energy demands that the plasma luminosity be less than the 
power injected at the base of the wind. 
This consideration -- together with ROSAT observations of the region \cite{Snowden1997, Almy2000} (which reveal a diffuse, $\lsim 10^7$ K plasma coincident with the bubbles \cite{Finkbeiner2004,Su2010} at relatively lower Galactic latitudes) --
imply that the plasma is extremely hot (as previously contemplated \cite{Su2010}): $> 3 \times 10^6$ K 
and of density (0.004--0.006) cm$^{-3}$.
This temperature scale is similar to that observed for `super-winds'  from the nuclei of
 star-forming galaxies \cite{Strickland2000}. 

Two effects may operate to ensure the plasma's high temperature.
 Firstly, it is  {\it injected} at very high temperature ($10^{7-8}$K),
 implying long radiative cooling times, $\sim$ Gyr
(there have been persistent, though disputed \cite{Revnivtsev2009}, claims that observations with Chandra \cite{Muno2004}, SUZAKU \cite{Koyama2007}, and previous X-ray instruments \cite{Yamauchi1990} suggest a $(6-9) \times10^7$ K, diffuse plasma in the $\sim100$ pc around the GC).
Secondly, the plasma may be (re-)heated by thermalization of initial bulk motion: the kinetic energy of each thermal proton in an 800 km/s outflow  corresponds to a temperature of $2 \times 10^7$ K. 
Moreover, given the  long formation timescales, coloumbic processes (acting on timescales $\ll 10^7$ yr) may distribute this final thermal energy between plasma protons and electrons.
%

A hadronic $\gamma$-ray scenario predicts the injection of relativisitic secondary electrons (and positrons) throughout the bubbles, synchrotron radiating on ambient magnetic fields.
In fact,
in the calorimetric limit,
 the kinematics of the charged and neutral pion decay chains
 (and the $\propto E_e^2$ dependence of the synchrotron critical frequency),  imply that $\nu L_\nu (synch)$
 is a significant fraction -- up to a quarter (depending on synchrotron vs. non-synchrotron energy loss rates) -- of
  $\nu  L_\nu(\pi^0)$. 
 For the best-fit $\gamma$-ray spectrum (fig.~\ref{fig_plotBubbleSpectrum}) we find a 20-60 GHz secondary electron synchrotron luminosity of $\sim 2 \times 10^{-36}$ erg/s for $B > 10^{-5}$ G (implying near energy density equipartition between the bubbles' magnetic field, CRs, and plasma).
 %
 %
 This is consistent with observations: the
 luminosity of the haze over the same WMAP bands (20--60 GHz) is $(1-5) \times 10^{36}$ erg/s \cite{Finkbeiner2004} to be contrasted with the $\sim4 \times 10^{37}$ erg/s in $\gamma$-rays \cite{Su2010}.
The energy-independent transport timescale associated with the putative wind (cf. \cite{Biermann2010}) can explain the hard spectrum of the microwave haze
(where it is measured inward of $\sim 18^\circ$ \cite{McQuinn2010}).
If it transpires that the microwave spectrum is hard out to the full, $\sim$ 10 kpc extent of the $\gamma$-ray bubbles this would be unexplained within our scenario and imply an additional electron population (injected or accelerated in situ as previously discussed).

Our scenario requires CR proton trapping for multi-Gyr timeframes.
Diffusive confinement could achieve this alone for a diffusion coefficient 1--2 orders of magnitude {\it smaller} than the Galactic plane value (cf. the electron/IC scenario).
This is reasonable: X-ray observations \cite{Yao2007} show that the plasma of the Galactic bulge is extremely turbulent. 
%
Moreover, a search for polarized emission in the WMAP data \cite{Gold2010} has yielded a negative result, reconcilable with this emission's being due to synchrotron if the magnetic field structure is highly tangled \cite{McQuinn2010}.
If the claimed \cite{Su2010} sharpness of the bubble edges is correct, it is likely that a magnetic draping effect \cite{Dursi2008} around the super-Alfvenic GC wind also helps trap the protons as apparently occurs in fossil cluster radio bubbles \cite{Ruszkowski2008}. 

The GC wind probably also has an important role in transporting positrons from the Galactic core out
into the Galactic bulge which is traced by 
the electron-positron annihilation line emission \cite{Knodlseder2005}.
This naturally predicts 511 keV emission out to $\sim$ 1 kpc from the GC, as observed, 
despite the fact that positrons must be injected into the ISM at energies of only a few MeV \cite{Agaronyan1981,Beacom2006}.

Finally, we predict there should be extended, $\sim$ TeV $\gamma$-radiation surrounding the Galactic nucleus on similar size scales to the bubbles with an intensity  $\leq E_\gamma^2 F_\gamma ({\rm TeV}) \sim 10^{-9}$ TeV cm$^{-1}$ s$^{-1}$ sr$^{-1}$ which should make an interesting target for 
future $\gamma$-ray studies.
Likewise, the region is a promising source for a future, Northern Hemisphere, km$^3$ neutrino telescope for which we estimate (assuming a $\gamma = 2.0$ proton spectrum, cut-off at 1 PeV)  $\sim$40 signal events above 10 TeV per annum (vs. $\sim$100 background events implying a $ 4\sigma$ detection in one year).
Of course, steepening or a lower-energy cut-off in the intrinsic proton spectrum or, indeed, loss of confinement at higher proton energies would reduce the significance of the bubbles as either TeV $\gamma$-ray or neutrino sources.


RMC is the recipient of an IIF Marie Curie Fellowship. The authors acknowledge conversation or correspondence with Joss Bland-Hawthorn, Valenti Bosch-Ramon,  Sabrina Casanova, Roger Clay,  David Jones, Casey Law, Mitya Khangulyan, Fulvio Melia, Brian Reville, Frank Rieger,  Ary Rodr{\'{\i}}guez-Gonz{\'a}lez, and  Heinz V{\"o}lk.



\end{document}